\documentclass{article}  

\begin{document}  
  
\title{Coulomb correlation and magnetic ordering in double-layered 
manganites: LaSr$_2$Mn$_2$O$_7$}  
  
\author{Julia E. Medvedeva\(^{1,}\)\(^{2,}\)\thanks{Corresponding author.
Fax: +1-847-491-5082. {\it E-mail address:} jem@pluto.phys.nwu.edu} ,  
Vladimir I. Anisimov\(^{2}\), Michael A. Korotin\(^{2}\),\\  
Oleg N. Mryasov\(^{1}\), Arthur J. Freeman\(^{1}\) \\  
\(^1\) Department  
of Physics and Astronomy, Northwestern University, \\ 
Evanston, IL 60208-3112 USA \\  
\(^2\) Institute of Metal Physics, Yekaterinburg 620219, Russia }  
  
\maketitle  
  
\begin{abstract}  
A detailed study of the electronic structure and magnetic configurations
of the 50 \% hole-doped double layered manganite LaSr$_2$Mn$_2$O$_7$ 
is presented. We demonstrate that the on-site Coulomb correlation (U) 
of Mn d electrons {\it (i)} significantly modifies the electronic 
structure, magnetic ordering (from FM to AFM), and interlayer exchange 
interactions, and {\it (ii)} promotes strong anisotropy in electrical transport,
reducing the effective hopping parameter along the {\it c} axis 
for electrically active $e_g$ electrons. This findng is consistent with
observations of anisotropic transport -- a property which sets this
manganite apart from conventional 3D systems.
A half-metallic band structure is predicted with both
the LSDA and LSDA+U methods. 
The experimentally observed A-type AFM ordering in LaSr$_2$Mn$_2$O$_7$
is found to be energetically more favorable with U $\geq$ 7 eV.
A simple interpretation of interlayer exchange coupling  
is given within double and super-exchange mechanisms based on the 
dependencies on U of the effective exchange parameters and $e_g$ state 
sub-band widths. 
\end{abstract}  
 

%\newpage

\section{Introduction}
  
The double layered CMR manganite materials, La$_{2-2x}$Sr$_{1+2x}$Mn$_2$O$_7$
\cite{Moritomo96},
demonstrate magnetic behaviour that is distinct from the perovskite manganites, 
La$_{1-x}$Sr$_x$MnO$_3$, including:  
(i) strong anisotropy of electrical (magneto-) transport 
\cite{Moritomo96,Kimura96,Perring97,Kimura98}; 
(ii) nearly 2D character of its magnetism and strong AFM magnetic correlations 
above T$_c$ \cite{mag:corr:98}; and 
(iii) anomalous magnetoelastic properties \cite{Kimura98}.  
  
Most experimental reports on the layered manganites have  
concentrated on the La$_{2-2x}$Sr$_{1+2x}$Mn$_2$O$_7$ compounds  
with $x\approx 0.4$,   
which are metallic and demonstrate strong magnetoresistive effects.   
Stochiometric LaSr$_2$Mn$_2$O$_7$ (x=0.5) exhibits interesting spin, charge 
and orbital ordering, and its resistivity is of the order of $~10^0 \, \Omega cm$
with a rather flat temperature dependence \cite{Kubota99}. 
This can be compared with  
the situation in perovskite manganites where undoped LaMnO$_3$ is an   
antiferromagnetic insulator with strong orbital ordering and   
La$_{1-x}$Sr$_x$MnO$_3$ with $x\approx 0.4$ is a metal with strong  
magnetoresistive effects \cite{Urushibara95}.  
  
In comparison with the pseudocubic perovskites with three-dimensional  
networks of MnO$_6$ octahedra (i.e., (La,Sr)MnO$_3$), these layered structures \\ 
(La,Sr)$_3$Mn$_2$O$_7$ have a reduced exchange coupling between the Mn ions 
along the {\it c} direction. Indeed, the pseudo-cubic perovskites show 
ferromagnetism and metallic conductivity over a wide range of hole doping,  
suggesting that the double-exchange mechanism is dominant among itinerant  
{\it e$_g$} electrons. In the double-layered case, with   
the two-dimensional Mn-O network, consisting of two perovskite blocks
separated by an intervening insulating layer of (La,Sr)O ions along the {\it c}
axis, the balance between antiferromagnetism and ferromagnetism is very 
sensitive to $e_g$ band filling \cite{Battle97}.  
Bilayer La$_{2-2x}$Sr$_{1+2x}$Mn$_2$O$_7$ demonstrates ferromagnetism 
in a doping range $x<0.39$, a canted antiferromagnetic structure 
for a hole concentration $0.39<x<0.48$, and layered antiferromagnetic
states for $x>0.48$ \cite{Kubota99a,Moritomo98}. 

From experimental data 
\cite{Kubota99,Battle97,Moritomo98,Hayashi98,Argyriou00,Chatterji00}, 
it is known that the ground state spin structure of LaSr$_2$Mn$_2$O$_7$ 
phase is A-type layered antiferromagnet (AFM), 
where the magnetic moments lie in the $ab$ plane and  
couple ferromagnetically within the single MnO$_2$ layer, but show  
AFM order between the respective MnO$_2$ layers within the bilayer unit.    
The interlayer coupling in the bilayer stack is ferromagnetic.  
   
One of the most striking properties of the bilayered LaSr$_2$Mn$_2$O$_7$ is  
the competition between the (charge-disordered) A-type AFM spin ordering  
with a Ne\'{e}l temperature T$_N \approx $170 K and the CE-type \cite{Wollan55} 
charge/orbital ordering which exists between $T_N$ and T$_{CO}$=210 K 
\cite{Chatterji00}. 
With decreasing temperature, the development of the charge/orbital ordering phase
is disrupted by the onset of the A-type AFM ordering at T$_N$ -- which may be
qualitatively explained by the onset of ferromagnetic order in each $ab$ plane.
 
An important issue in the theory of CMR oxides is the role of Coulomb 
correlation. An accurate treatment of correlation may significantly affect 
the balance between ferromagnetic and antiferromagnetic interactions 
and hence the magnetic ground state. LDA band structure calculations 
for CMR perovskites demonstrated the possibility of LDA theory predicting   
the correct magnetic ground state \cite{Pickett96}.
For the double layered perovskites, an investigation of the reliability of this 
theory has not yet been performed. In their report on the electronic 
structure calculations of LaSr$_2$Mn$_2$O$_7$ within the general 
gradient approximation (GGA) \cite{Boer99}, Boer and de Groot suggested to take 
into account some additional correlation corrections to obtain 
a truly half-metallic electronic structure in the layered manganites.
There is one report of an LDA+U calculation \cite{Dessau98} performed  
for the doping level of 0.4 holes per Mn site  
(La$_{1.2}$Sr$_{1.8}$Mn$_2$O$_7$), but details of their calculation 
have not been presented and the important question of how correlations 
affect the magnetic ground state has not been investigated. 
  
The purpose of our work is to study the electronic structure which governs 
the microscopic origin of the magnetic phenomena in the bilayered manganite, 
and to determine its dependence on Coulomb correlations and its relation   
with magnetic properties.  
We consider the difference in the total energies, exchange interaction 
parameters, the population of Mn-d and O-p states near the Fermi level 
and d sub-band widths for ferromagnetic (FM) and A-type antiferromagnetic 
(AFM) configurations as a function of the Coulomb correlation 
parameter U and give a simple interpretation of exchange coupling between
Mn layers in LaSr$_2$Mn$_2$O$_7$ within the double exchange model.
In contrast to the case of LaMnO$_3$, we find that the inclusion of U
in LaSr$_2$Mn$_2$O$_7$ results in important differences in the $e_g$-band 
characteristics that are responsible for the observed FM instability
and for strong anisotropy in the electrical transport.  

\section{Methodology}
  
The LSDA and LSDA+U calculations were realized in the frame-work of the   
linear-muffin-tin-orbital method in the atomic sphere approximation  
(LMTO-ASA) \cite{Andersen75}. We used the von Barth-Hedin-Janak form 
\cite{Barth75} for exchange-correlation potential.  
The crystal parameters were taken from Ref. \cite{Argyriou00}.  
The atomic sphere radii were: R(La(Sr))=3.4 a.u., R(Mn)=2.8 a.u., and  
R(O)=2.2 a.u. From the constrained LSDA supercell calculations 
\cite{Gunnarsson89,Anisimov91},  
we obtained values of the Coulomb and exchange parameters to be
U=7.2 eV and J=0.78 eV. These values are typical for the transition-metal 
oxides \cite{ldau,Satpathy96,Solovyev96}.    

\section{Results and Discussion}
\subsection{Electronic structure}
  
As a first step, we calculated the electronic structure  
of FM (all atoms in every layer and between layers are ordered 
ferromagnetically) and A-type AFM (ferromagnetic layers stacked 
antiferromagnetically) LaSr$_2$Mn$_2$O$_7$ by the standard LSDA method (U=0).
The total and projected densities of states (DOS) for these cases are shown 
in Fig. 1. For the majority-spin channel (solid line), 
Mn 3d states form the bands between 2.2 eV below E$_F$ and 2.5 eV 
above E$_F$. As seen from the figure, the $e_g$ ($x^2-y^2$ 
and $3z^2-r^2$) bands are partially filled, cross E$_F$, and are rather 
broad compared to the $t_{2g}$ ($xy$ and degenerate $xz$, $yz$) bands
which are about 1.2 eV wide and lie 1 eV below E$_F$. 
The exchange interaction splits 
the Mn 3d states such that the $t_{2g}$ and $e_g$ minority-spin bands 
are located 2.6 eV higher in energy than the  majority-spin states.  
We obtained a band gap of 2.5 eV in the minority-spin $d$ bands  
and a band structure that is in a very good agreement with that obtained
by the full-potential LMTO method \cite{Huang00} (but 
their band gap within the minority spin channel is about 1.7-1.9 eV). 
Our results for the ferromagnetic state with U=0 differ from the results 
of Dessau {\it et al} \cite{Dessau98} where a U of 2 eV was used to obtain 
a gap at E$_F$ in the minority-spin bands for La$_{1.2}$Sr$_{1.8}$Mn$_2$O$_7$.  
It should be noted, that the Fermi level in our calculations was found  
to lie at the bottom of the $t_{2g}$ minority conduction band 
(as in the FLMTO calculation \cite{Huang00}), so the electronic structure 
just misses to being a half-metallic one.  
    
In order to investigate the role
of on-site Coulomb correlation, we calculated 
the electronic structures of LaSr$_2$Mn$_2$O$_7$ for both FM and A-type 
AFM orderings using a number of values of  
U, namely 2, 4, 6, 7.2, 8, 9 and 10 eV. The total and projected DOS's of 
the FM and AFM cases with U=7.2 eV are shown in Fig. 2.
Only majority $e_g$ states make a significant contribution to the DOS at E$_F$.  
For U=0, one can expect three-dimensional conduction in both spin 
channels (Fig. 1), while for U$>$0 we obtain a half-metallic state 
-- electron conduction is possible only within majority spin sublattice.
We want to emphasize here analogues in the main band characteristics near 
E$_F$ in the calculated electronic spectra of double layered  
and doped perovskite manganites \cite{Solovyev00}. 
And since doped perovskite manganites are found to be half-metallic 
\cite{Wei97,Park98}, this may also be the case for the double-layered manganite.

\subsection{Anisotropy in electrical transport}

To investigate the variation (behavior) of the electrically active states as a function of 
on-site Coulomb correlation, we plot in Fig. 3 the dependence on U of the $3z^2-r^2$ 
and $x^2-y^2$ states contribution to the DOS at E$_F$ (Fig. 3(a))
and to the number of states integrated in a small energy window of 0.3 eV 
just above E$_F$ (denoted as N$_{3z^2-r^2}$ and N$_{x^2-y^2}$ in Fig. 3(b)). 
The difference between FM and AFM phases for a particular U value is clearly 
seen from Fig. 3(a). For the FM phase (dotted lines), both contributions of the 
$3z^2-r^2$ (circles) and $x^2-y^2$ (crosses) states decrease with U in almost 
the same way, and only large U values (U$>$7 eV) result in a splitting 
of the $e_g$ states. 
The most significant changes occur for AFM spin alignment (solid lines).
The $e_g$ states are already split in the LSDA calculation (U=0)
and the difference between the electron population of
$3z^2-r^2$ and $x^2-y^2$ orbitals is clearly seen (Fig. 3(a)).
Further, while the contribution of the $x^2-y^2$ state to the DOS at E$_F$ 
does not change with an increase of U, the ${3z^2-r^2}$ contribution, which is 
almost two times bigger than that of ${x^2-y^2}$ for U=0, drastically decreases 
with U, taking its minimum for U$>$7 eV.

As can be seen from Fig. 3(b), the situation does not change qualitativly 
in the energy window just above E$_F$, encouraging us to draw the
following conclusions:
For small U values we have 3D conduction, while for U$>$6
the electron conduction (hopping) along the $c$ axis becomes very small.
Thus, on-site correlation treated within LSDA+U promotes 2D type 
(in-plane) electronic behavior in the system.
We need to add here that the significant difference between 
$3z^2-r^2$ and $x^2-y^2$ states (due to an decrease of the band width 
as well as a shift to higher energy of the $3z^2-r^2$ state with increase of U) 
is an essential feature of LaSr$_2$Mn$_2$O$_7$.
It sets the double layered manganite apart from perovskite manganites
(i.e., (La,Sr)MnO$_3$) in which the main energy characteristics 
such as widths and centers of both $e_g$ bands are very close.

The anisotropy in the population of the two $e_g$ states established above 
can be illustrated more clearly in the plot of the angular 
distribution of the electron density in the 0.3 eV energy window 
above E$_F$ for U=0 and for the calculated value, U=7 eV (c.f., Fig. 4).
A comparison of the lowest spin configurations (FM for U=0 (Fig. 4(a)) 
and AFM for U=7 eV (Fig. 4(d)) as obtained from total energy differences)
shows that both on-site correlation and change in spin ordering
significantly decrease the number of electrons that contribute to electrical
transport along the $c$ axis. This is contrary to the in-plane state 
which remain less sensitive to the spin alignment change and influence of U.
(Note, however, that only on-site correlation supresses the minority $t_{2g}$ 
($xy$) state  contribution to the electron density above E$_F$, 
resulting in a half-metallic state, as mentioned above.)

The decisive role of the on-site Coulomb correlation in 
lowering the dimensionality of the conduction band is clearly seen from a
comparison of two AFM cases with U=0 and U=7 eV (Fig. 4(b) and 4(d)):
for U=7 eV, only the $x^2-y^2$ orbital has some contribution just above 
E$_F$ and hopping along the $c$ axis is negligible (the $3z^2-r^2$ state 
is responsible for it). 
This is consistent with the point that the main issue for realizing 
the layered AFM state is the strong anisotropy in the population of the 
two-dimensional $d_{x^2-y^2}$ state and the one-dimensional 
$d_{3z^2-r^2}$ state.
For the FM state, the orbital configurations have more 
electron population freedom, because three crystallographic axes 
are equivalent. Qualitatively this situation does not change as 
U increases.
According to the qualitative picture of the simple double exchange model,
an electrically active electron has a finite hopping probability (t)
between ferromagnetically ordered Mn ions, but this hopping vanishes
in the case of antiferromagnetic spin alignment due to strong Hund coupling 
(t$<<$J). Figure 4(b) clearly shows the failure of the LSDA sheme.
Thus, taking into account on-site Coloumb correlation gives us the correct
picture of 2D-like charge transport in the layered manganite: the increase of U
suppresses the $c$ axis transport, keeping electron conduction in the $ab$ plane.
However, it should be noted that the transition from FM to AFM ordering 
(Fig. 4(c) and 4(d)) also contributes to this tendency, but to a lesser degree 
than on-site correlation -- as can be seen from a comparison of Fig. 4(a) and 4(d).

\subsection{Magnetic ordering}

To investigate how the magnetic interaction and ground state depend on U,
we calculated total energy differences for U=0, 2, 4, 6, 7.2, 8, 9 
and 10 eV. The results are shown in Fig. 5(a), where the total energy 
of the FM spin ordering for each value of U is taken to be zero. 
Although the difference of the two calculated spin alignments is  
$\leq$ 0.1 eV for all cases (and so we have almost degenerate magnetic 
states), for U=0, U=2 and U=4 eV the FM ordering is preferred, while  
for U=6, 7.2, 8, 9 and 10 eV the AFM phase has lower energy.  
Thus, the on-site Mn-d electron Coulomb correlations modify 
the magnetic ordering from FM to A-type AFM -- as experimentally observed. 
Note, that the total energy difference between FM and AFM states 
does not change for U $\geq$ 8 eV.
To determine the character of the exchange coupling in the layered manganite
and to understand why the experimentally observed A-type AFM spin ordering
is simulated only with U$>$7 eV, we also calculated the effective exchange 
interaction parameters (for computational details see \cite{lichtan}).
In Fig. 5(b), the values of the d-d effective exchange parameters, 
$J_{dd}$, between nearest Mn neighbours which belong to the different 
layers of one bilayer are shown as a function of U.
They are seen to follow the behavior of the total energy differences 
shown in Fig. 5(a).  

Usually, the exchange coupling is described within the superexchange (SEX) 
and double exchange (DEX) models and corresponding solutions of the simple
Hamiltonian problem with the following parameters: hopping between Mn ions (t),
J (DEX model); and t, U (SEX model). To provide interpretation to this model, 
we plot the $e_g$-sub-band widths of the FM phase for a number of U values 
in Fig. 5(c) (denoted as W$_{3z^2-r^2}$ and W$_{x^2-y^2}$ below). 
As can be seen, the $x^2-y^2$ band width almost 
does not change its value with U, and the main differences in the electronic 
spectra due to an increase of U are connected with the $3z^2-r^2$ state.
The decrease of the difference between the W$_{3z^2-r^2}$ and W$_{x^2-y^2}$ 
band widths from U=0 up to U$~$5 eV in Fig. 5(c), coming from
the increase of W$_{3z^2-r^2}$, promotes the DEX interactions 
mediated by itinerant electrons, Fig. 5(b).
Further, the positive (FM) contribution to the exchange interaction changes 
its sign at about U=7 eV, where a sharp decrease of the W$_{3z^2-r^2}$  
value for the FM phase occurs. In addition, due to a redistribution of 
the electron density between states, W$_{x^2-y^2}$ increases slightly,
and so the splitting of these two $e_g$-sub-bands increases.
Thus, Coulomb correlations supress the DEX contributions 
to the interlayer exchange interaction energy, and we believe that the
sharp $3z^2-r^2$ band width narrowing in the range of 6$<$U$<$7 eV
is responsible for the FM instability, and hence, is the mechanism 
that produces the change in magnetic ordering. In addition, this splitting
of the $e_g$ states promotes strong anisotropy in the electrical transport,
as described above.

\section{Conclusion}

In summary, we have calculated the electronic structure of double layered \\
LaSr$_2$Mn$_2$O$_7$ using both the LSDA and LSDA+U methods and have 
investigated the influence of the Coulomb interaction parameter 
on both the electronic structure and the magnetic ordering.  
The main factors governing band formation in LaSr$_2$Mn$_2$O$_7$
are: (i) the exchange splitting with almost unoccupied minority 
spin Mn-3d states; (ii) the ligand field splitting of the $t_{2g}$ and 
$e_g$ states; and (iii) futher splitting of the $e_g$ states
with increase of U. We consider (iii) an essential factor 
in the layered manganites. We have examined FM and A-type AFM
ground states of this compound by comparing their total energy 
differences for a number of U values (from 0 to 10 eV) and found that 
the experimentally observed magnetic ordering is reproduced within 
LSDA+U only for U$\geq$7 eV - in contrast with the 3D manganites 
(LaMnO$_3$) where LSDA gives the true magnetic ground state.
It was shown that the correlations reduce significantly the effective 
hopping parameter for the electrically active $e_g$ electrons along 
the $c$ axis, so that taking into account the Coloumb 
correlation gives the correct 2D-picture in the layered 
manganite. The strong anisotropy in electrical transport which we obtained
is consistent with observations of anisotropic transport - 
a property which sets this manganite apart from conventional 3D systems.

\vspace{1cm}
 
Work at Northwestern University supported by the U.S. Department of Energy
(grant No. DE-F602-88ER45372).

\newpage 
 
%%%%%%%%%%%%%%%%%%%%%%%%%%%%%%%%%%%%%%%%%%%%%%%%%%%%%%%%%%  
%   F I G U R E S  
%%%%%%%%%%%%%%%%%%%%%%%%%%%%%%%%%%%%%%%%%%%%%%%%%%%%%%%%%%  
% 1.  
\begin{figure}[t]  
\caption{  
Calculated LSDA total and projected densities of states (in states/eV-unit cell)
of the ferromagnetic (thin line) and antiferromagnetic (thick line) phases.   
The Fermi level is located at 0 eV.}  
\end{figure}  
  
% 2.  
\begin{figure}[t]  
\caption{  
The total and projected densities of states (in states/eV-unit cell) 
of the ferromagnetic (thin line) and antiferromagnetic (rich line) 
phases from the LSDA+U calculation, U=7.2 eV.   
The Fermi level is located at 0 eV.}  
\end{figure}  
  
% 3.  
\begin{figure}  
\caption{  
Contribution of $3z^2-r^2$ (circles) and $x^2-y^2$ (crosses) states to 
(a) the density of states at E$_F$ and (b) the number of states (NOS) 
integrated in a small energy window of 0.3 eV just above E$_F$ 
as a function of Coulomb U.
Solid and dashed lines correspond to AFM and FM cases, 
respectivly.  }  
\end{figure}   
  
% 4.  
\begin{figure}  
\caption{
The calculated angular electron density distribution for the energy window of
0.3 eV width just above E$_F$ for (a) FM, U=0; (b) A-type AFM, U=0;
(c) FM, U=7.2 eV and (d) A-type AFM, U=7.2 eV.
Only one MnO$_6$ octahedron which belongs to the upper layer of the bilayer 
is shown.
Solid line is for majority spin and dashed line is for minority spin. }  
\end{figure}   
  
% 5.  
\begin{figure}  
\caption{
The influence of U on
(a) the total energy difference between FM and A-type AFM phases;
(b) the calculated interlayer exchange interaction parameters 
for FM phase;
(c) the width of the $3z^2-r^2$ (circles) and $x^2-y^2$ (crosses)
bands for FM phase.}  
\end{figure}     
 
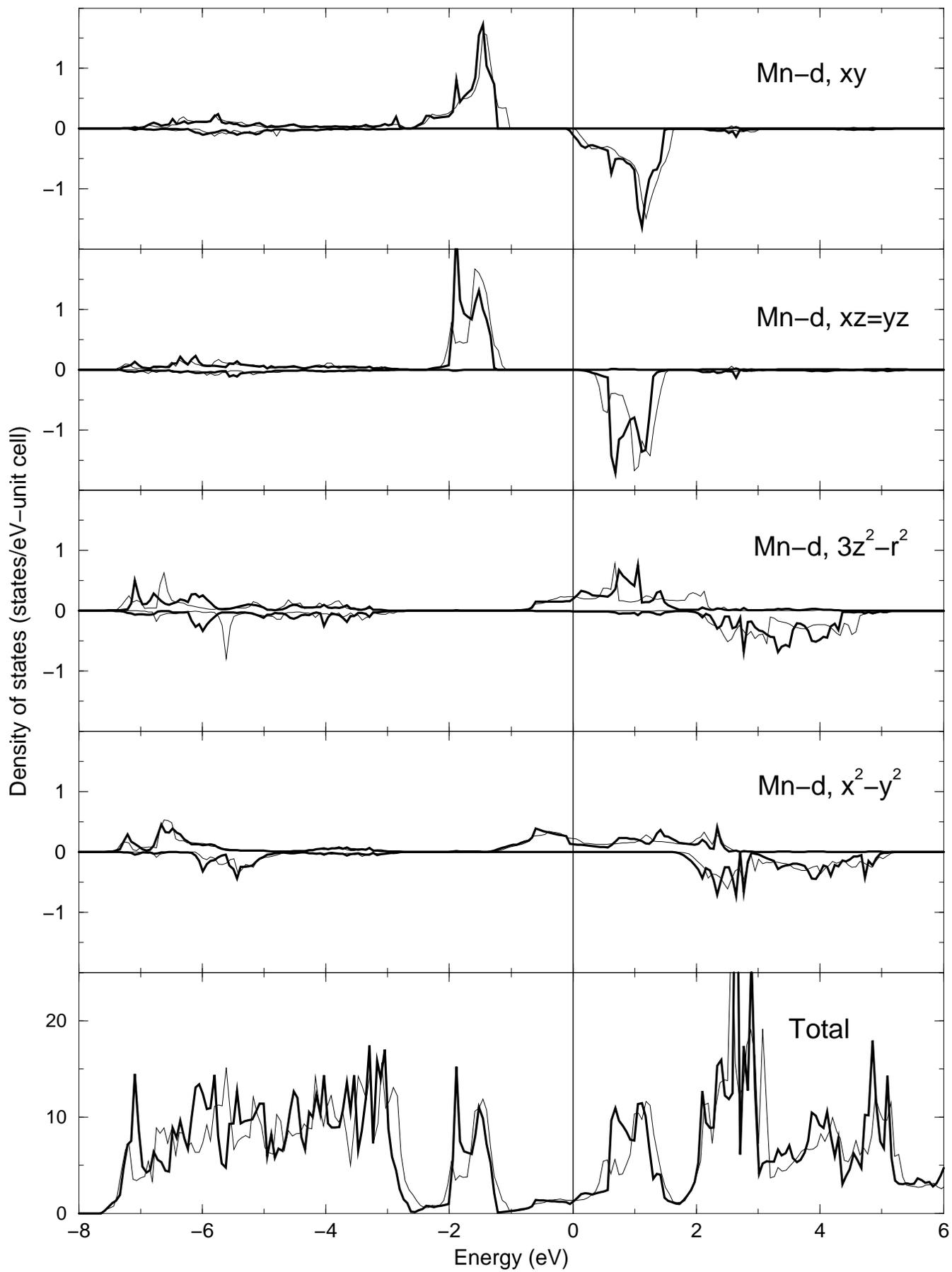
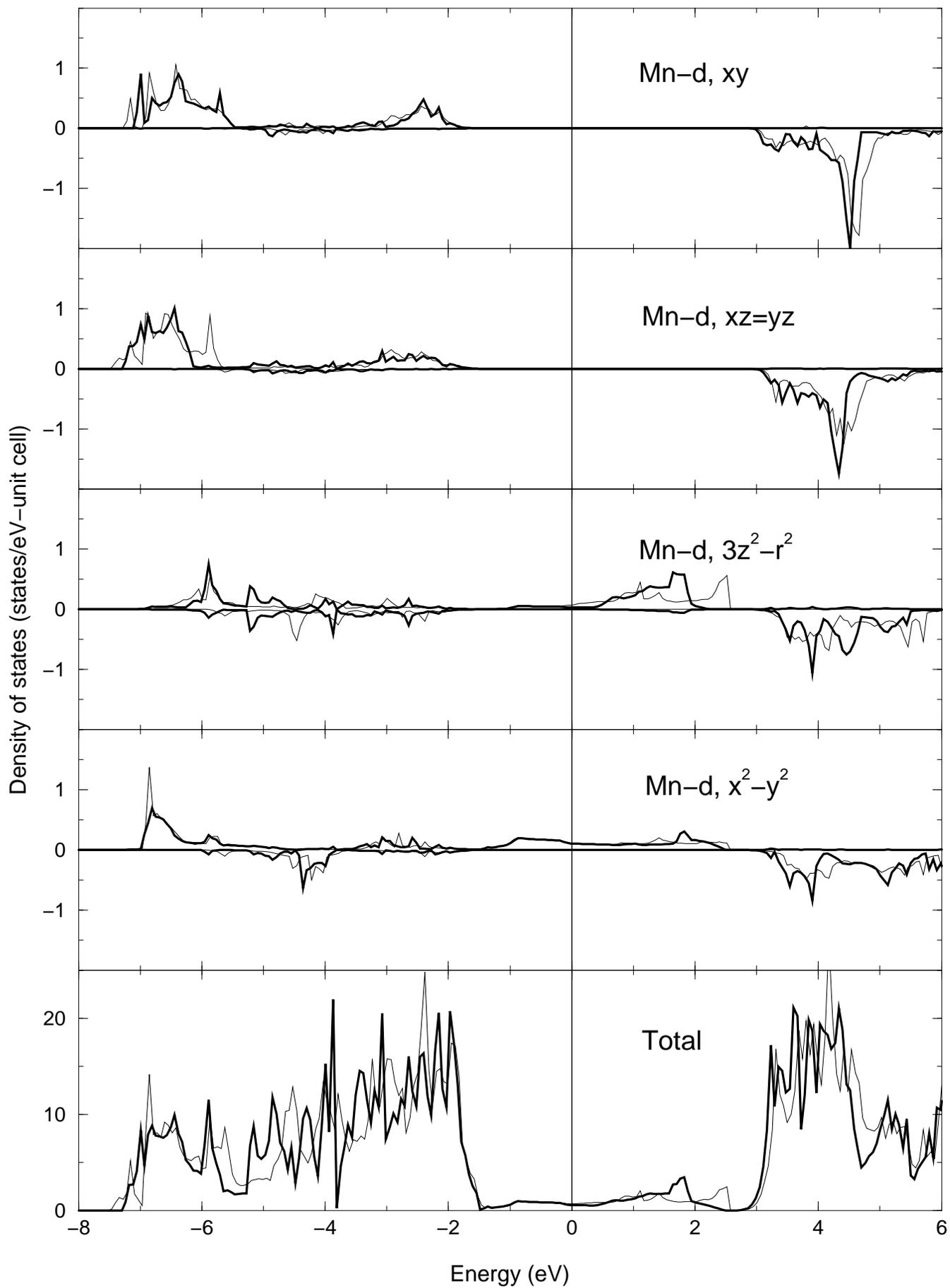
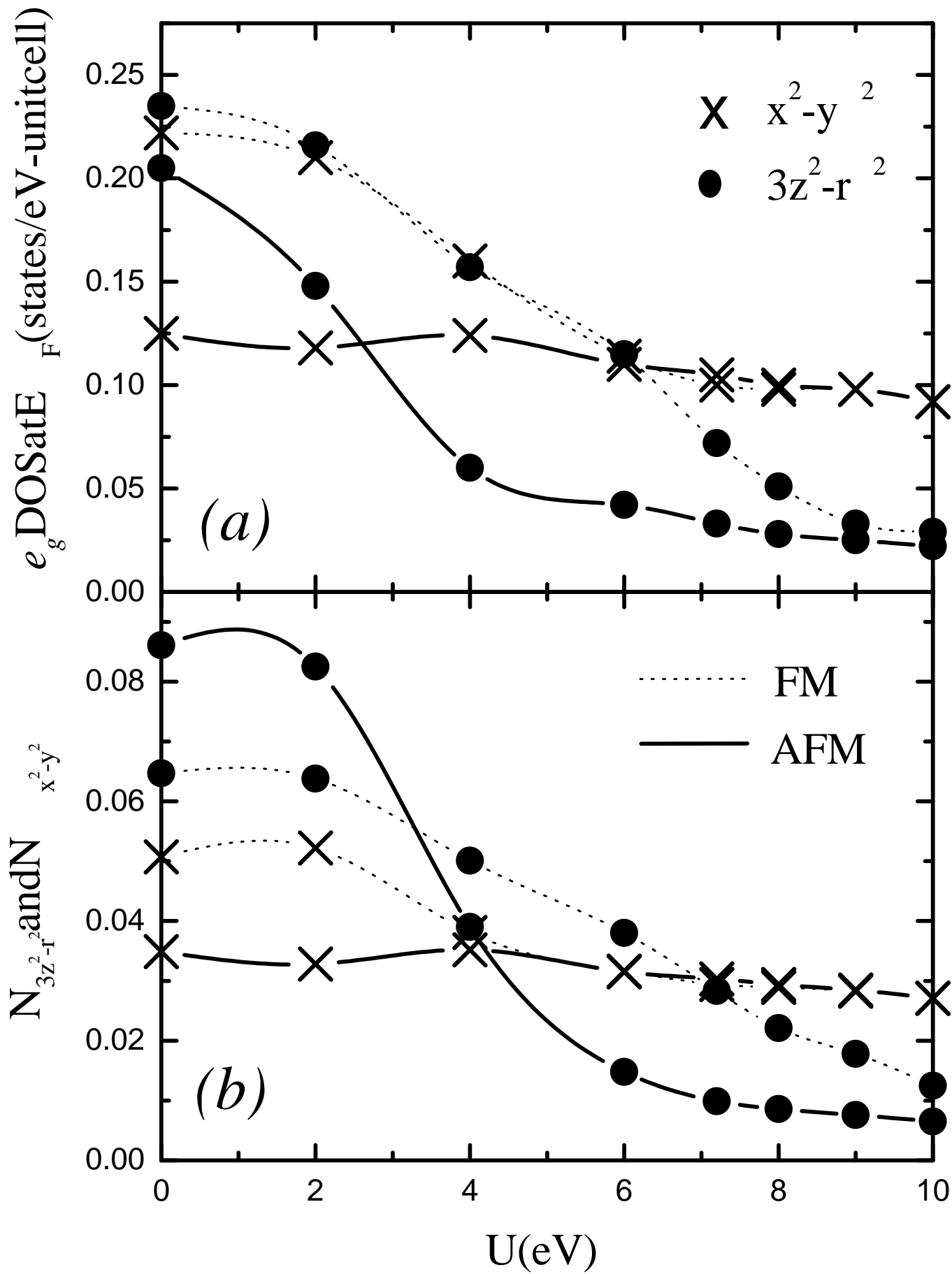
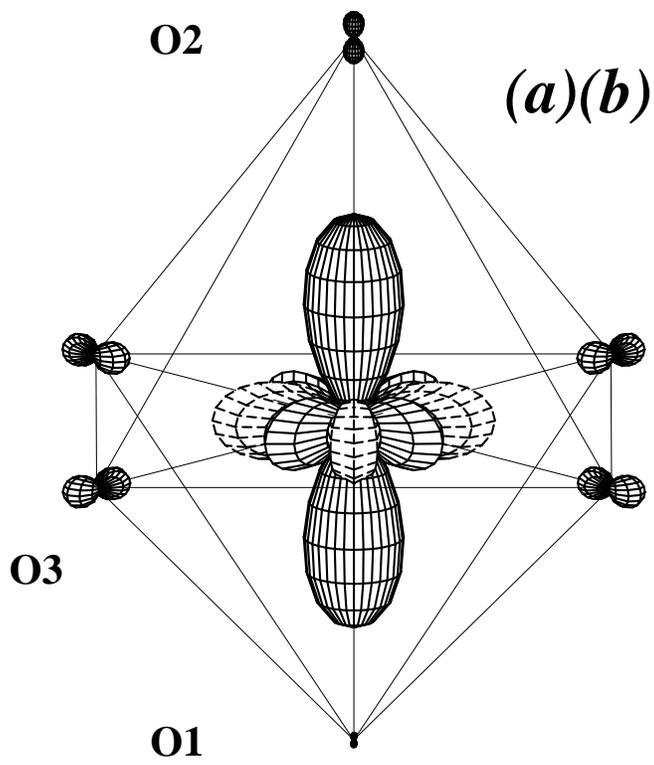
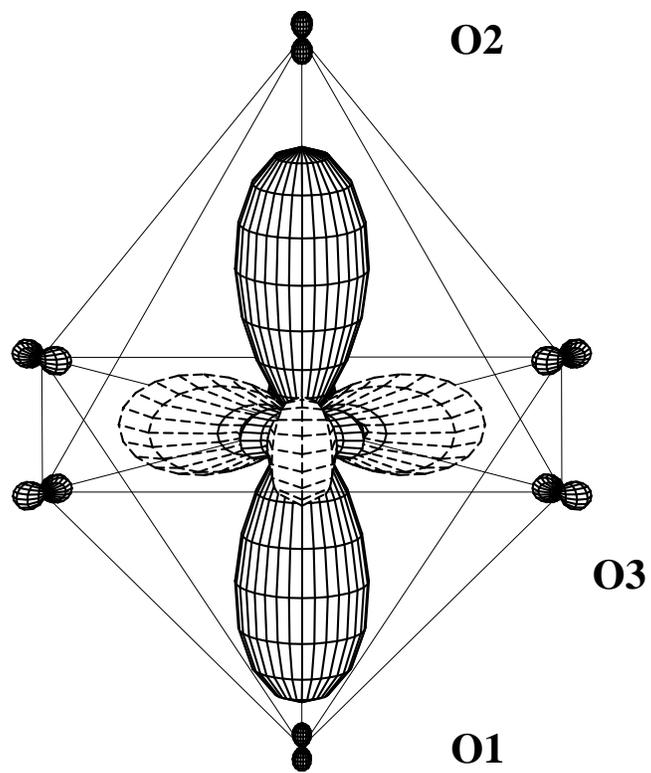
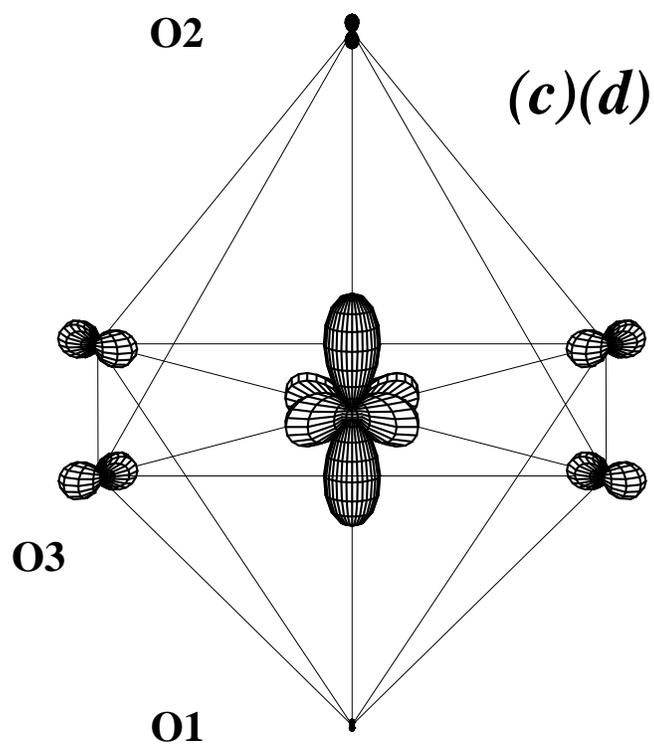
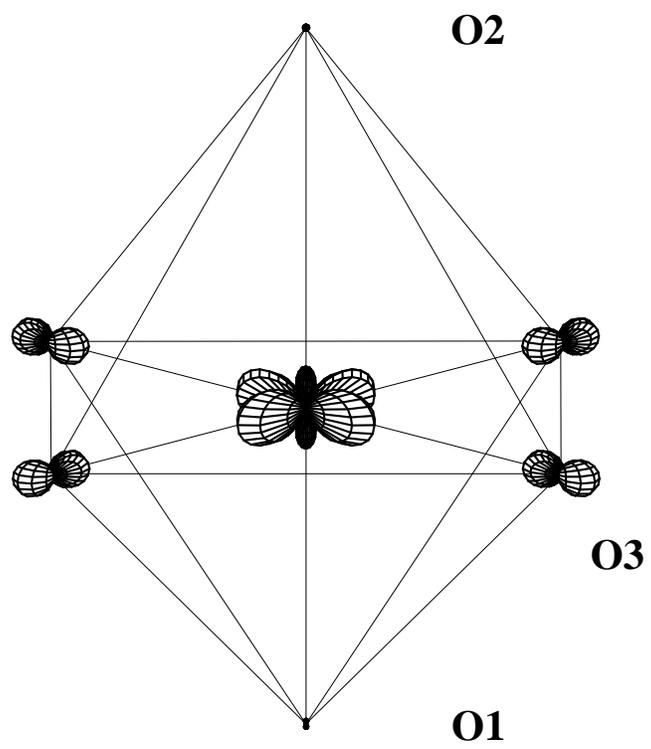
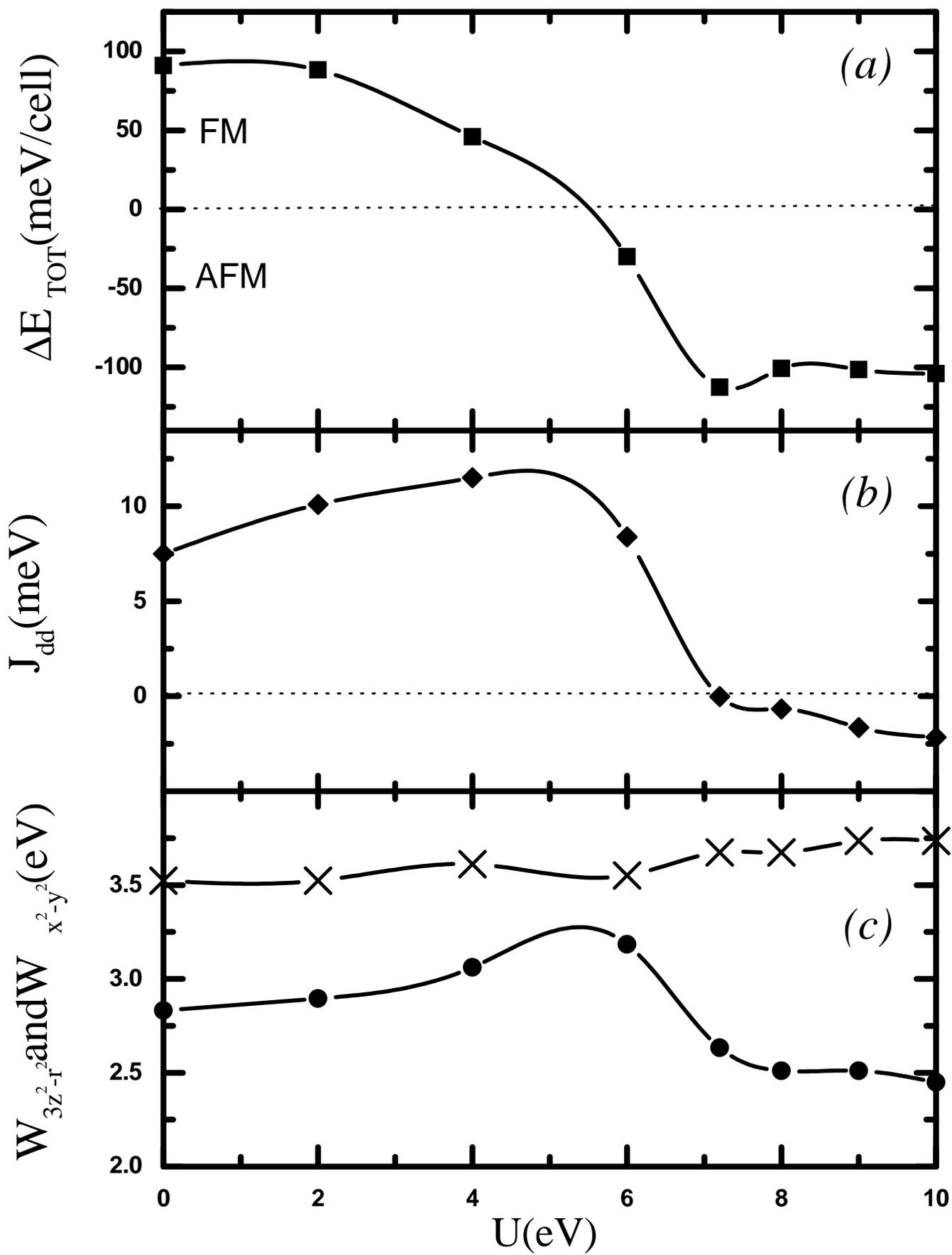
\end{document}